\documentclass[conference]{IEEEtran}
\IEEEoverridecommandlockouts
\usepackage{cite}
\usepackage{amsmath,amssymb,amsfonts}
\usepackage[numbers,sort&compress,sectionbib]{natbib}
\usepackage{algorithmic}
\usepackage{graphicx}
\usepackage{textcomp}
\usepackage{xcolor}
\usepackage{framed}
\def\BibTeX{{\rm B\kern-.05em{\sc i\kern-.025em b}\kern-.08em
    T\kern-.1667em\lower.7ex\hbox{E}\kern-.125emX}}
\newcommand{\footURL}[1]{\footnote{\url{#1}}}

\newcommand{\footReURL}[2]{\footnote{\label{#2}\url{#1}}}
\makeatletter
\newcommand\footnoteref[1]{\protected@xdef\@thefnmark{\ref{#1}}\@footnotemark}
\makeatother
\usepackage{subcaption}
\usepackage{ragged2e}
\usepackage{xfp}

\usepackage{hyperref}
 \hypersetup{
 colorlinks   = true,
 citecolor    = blue,
 urlcolor    =  blue
 }

\newcommand{\FR}{\textit{FRW}}
\newcommand{\FRW}{\FR{} data set}

\newcommand{\FRWFL}{\FR\textit{-FL}}

\newcommand{\dnd}{D\&D}

\newcommand{\articles}{over 47,800 articles}
\newcommand{\oldarticles}{over 45,200 articles}
\newcommand{\articleDate}{February 2023}

\newcommand{\yearsannotated}{2964}

\newcommand{\listoneannotations}{399}
\newcommand{\listtwoannotations}{242}
\newcommand{\typedannotations}{379}

\newcommand{\othersannotated}{%
  \fpeval{\listoneannotations+\listtwoannotations+\typedannotations}%
}
\newcommand{\annotationsdone}{%
  \fpeval{\othersannotated+\yearsannotated}%
}

\usepackage{listings}

\newcommand{\code}[1]{\texttt{#1}}

\newcommand{\shade}{\texttt{SHADE}}
    
\begin{document}

\title{SHADE: Semantic Hypernym Annotator for Domain-specific Entities - DnD Domain Use Case\\
%\thanks{\textbf{The APC was funded by   University of Moratuwa Senate Research Committee Fund for publications.
%This publication was funded by the publication support scheme of the SRC of University of Moratuwa (UoM).}}
}

\author{Akila Peiris, Nisansa de Silva \\
  Department of Computer Science \& Engineering,\\University of Moratuwa, Sri Lanka \\
  \texttt{\{akila.21,nisansadds\}@cse.mrt.ac.lk} \\}

% \author{Anonymous submission}

\maketitle

\begin{abstract}
Manual data annotation is an important NLP task but one that takes considerable amount of resources and effort. In spite of the costs, labeling and categorizing entities is essential for NLP tasks such as semantic evaluation. Even though annotation can be done by non-experts in most cases, due to the fact that this requires human labor, the process is costly. Another major challenge encountered in data annotation is maintaining the annotation consistency. Annotation efforts are typically carried out by teams of multiple annotators. The annotations need to maintain the consistency in relation to both the domain truth and annotation format while reducing human errors. Annotating a specialized domain that deviates significantly from the general domain, such as fantasy literature, will see a lot of human error and annotator disagreement. So it is vital that proper guidelines and error reduction mechanisms are enforced. One such way to enforce these constraints is using a specialized application. Such an app can ensure that the notations are consistent, and the labels can be pre-defined or restricted reducing the room for errors. In this paper, we present \shade, an annotation software that can be used to annotate entities in the high fantasy literature domain. Specifically in Dungeons and Dragons lore extracted from the Forgotten Realms Fandom Wiki.

\end{abstract}
\renewcommand\IEEEkeywordsname{Keywords}
\begin{IEEEkeywords}
data annotation, data extraction, natural language processing, fantasy literature, dungeons and dragons
\end{IEEEkeywords}

\section{Introduction}
Dungeons and Dragons (also known as \dnd{} or DnD) is a turn-based tabletop role-playing game which has gained immense popularity in the last 5 decades. In this game, typically set in a fantasy setting, a group of players role-play characters and go on adventures\footnote{\textit{Adventure} or \textit{adventure module} refers to a game guide that manages player knowledge and activities for a specific scenario typically with a cohesive narrative.} conducted by one player in the special role of \textit{Dungeon Master}. It is a form of an open-ended game, meaning that there are no \textit{correct} paths to play the game~\cite{squire2007open}. It is up to the player's interpretation of the game world (sandbox) and its underlying rules to progress in the game. In \dnd{}, the rules applicable for gameplay come from the official resources such as the \textit{Player's Handbook}~\cite{crawford2014player} and the \textit{Dungeon Master's Guide}~\cite{master2014dungeon} which are categorized as rule books. 
%They are also versionized to support the different \dnd{} versions with fifth edition (5e) being the latest and greatest. 
Apart from the main rule books, there can be campaign\footnote{\textit{Campaign} refers to a game guide for an overarching storyline across multiple adventures, typically with the same set of characters.} rules as well as non-official rules implemented by the \textit{Dungeon Master}.
These game world rules not only refer to the restrictions for the \textit{player actions} but it is also intertwined with the setting\footnote{The \textit{setting} generally is \textit{the world} of the game. In some instances, however, a setting may incorporate multiple worlds.}.

The setting includes the lore (history and current status of the setting), inhabitants, character classes, weapons, artifacts, magical spells, potions and many more to help the gameplay. 
Almost all of these come with their own statuses and other measurements. Compared to the rule books, the \textit{setting} includes most of the domain-specific named entities which may differ from the general domain.
% requiring domain-sepcific NLP solutions~\cite{sugathadasa2017synergistic}
Identifying the category of a given entity is an essential part of the gameplay in order to understand the semantics in relation to the game domain. For example, while Merriam-Webster online dictionary defines \textit{monstrosity}\footURL{https://www.merriam-webster.com/dictionary/monstrosity} and \textit{monster}\footReURL{https://www.merriam-webster.com/dictionary/monster}{MonsterWebRef} as synonyms, in \dnd{} they have a hierarchical relationship. In \dnd, \textit{Monster}\footReURL{https://bit.ly/DnDBmonster}{MonsterRef} is the semantic hypernym \cite{de2013semi} (the superclass) of \textit{Monstrosity}. 
Knowing the semantic relationship between the two is a major part of disambiguation rules. For example, a restriction or a benefit applicable to \textit{monstrosity} (sub-class) type does not necessarily apply to \textit{monster} (superclass) type. Although, a condition that is applicable to \textit{monster} (superclass), is typically applicable to \textit{monstrosity} (sub-class). Another close example would be that in \dnd{}, \textit{beasts} also fall under the category of \textit{monsters}\footnoteref{MonsterRef} even though in general domain beast would refer to an animal,  typically four footed\footURL{https://www.merriam-webster.com/dictionary/beast}, as opposed to a \textit{monster} where the defining trait would be the abnormality or the terrifying nature\footnoteref{MonsterWebRef}.

Data annotation which is typically done manually has the possibility of generating inconsistent annotations. Inconsistency in data annotation can occur in 2 ways. The first one is non-expert annotations. If the annotator is not an expert in the domain, the annotations may not be as accurate as one done by an expert. The second type of inconsistency can occur due to human error. Whether typos in the annotations, omissions of parts of the label text, or even unwanted additions to the label can all lead to inconsistency in annotations. The second type of error can easily occur when annotating data in the fantasy domain which has a considerable amount of deviation from the general domain. For example, there can be different spellings for words\footnote{For example, the linguistic arguments on the plural of \textit{dwarf}~\cite{dwarves2004liberman}}. There can also be an inordinate amount of accented words such as \textit{Faerûn}\footURL{https://forgottenrealms.fandom.com/wiki/Faerûn} as well as fantasy-esque words violating spelling norms (e.g.\textit{Eilistraee}\footURL{https://forgottenrealms.fandom.com/wiki/Eilistraee}). These factors should be considered when annotating such a data set.

% Nisansa is here (deep read)

We present \shade, a web-based annotation application to annotate entities with the category label in the \dnd{} domain. As the most popular out of all the \dnd{} settings, and the defacto default setting in \dnd{} 5e, the \textit{Forgotten Realms} setting has a vast collection of resources including the
Forgotten Realms Wikia\footReURL{https://forgottenrealms.fandom.com}{forgottenrealmswikiref} which has \articles as of \articleDate. Hence, we have chosen this as our lexical resource. \shade{} populates 2 different lists of potential labels for a given entity and is capable of capturing the annotation tag along with a weight depending on the source of the label on a 3 weight scale with 1 being the most important source (1: from links, 2: from noun phrases, 3: manually typed in). By limiting the manually typed in inputs and instead extracting the tags from the lexical resource itself, we can minimize most of the human errors we mentioned above.

Apart from the entity name and the multiple label tagging option, the UI also renders the formatted clean text version of the first paragraph of the wiki article associated with the entity being tagged. This way the non-expert annotators can get a context of the entity in question without being distracted by the formatting, miscellaneous information in the wiki page and the full article content. This is done to improve efficiency of the annotators and as a small remedy to bridge the gap between expert annotations and non-expert annotations.

\section{Related Work}

Domain specific data sets~\cite{rameshkumar2020storytelling, peiris2022synthesis} are useful in a number of NLP tasks such as text generation and abstractive summarization. Entity classification data sets can be used in semantic similarity comparison evaluations in text generation and abstractive summarization tasks where the generated text must adhere to semantics according to the given domain.

The biggest challenge in manual annotation is the cost and time required for the task. To address this, researchers have explored other avenues. One such alternative is the use of active learning techniques to reduce the amount of annotation required~\cite{olsson2009literature}. It involve selecting examples for annotation based on the current state of the classifier, with the goal of minimizing the overall annotation effort while still ensuring that the data is annotated accurately. The learning process takes in advice from the user on more complex queries.

Another alternative is crowd-sourcing~\cite{snow-etal-2008-cheap,dumitrache2015achieving}. Crowd-sourcing can offer several advantages, including increased annotation efficiency and scalability, as well as access to a wider range of annotators with different levels of expertise. However, it is important to consider the potential biases and limitations of crowd-sourced annotations, and to carefully design and evaluate the annotation process to ensure that the data is annotated accurately and consistently. \citet{snow-etal-2008-cheap} observes that it takes at least 4 non-expert annotations per item to match that of an expert level annotation.

\subsection{Annotation consistency}

When considering manual annotation which is the most common way of annotating data, whether it is via crowd sourcing or by an dedicated annotators, multiple annotators are needed to annotate a reasonably sized data set. There in lies one of the other biggest challenge encountered in manual data annotation; consistency of the annotations. 

Inter-annotator agreement (IAA), refers to the consistency of annotations produced by different annotators. Several studies have investigated the factors that affect IAA, such as annotator expertise, annotation guidelines, and annotation complexity~\cite{artstein2008inter,ouyang2017exploring}.The results of these studies suggest that providing clear and detailed annotation guidelines and ensuring that annotators have adequate training and expertise can improve IAA.

\subsection{\FRW}
The \FRW{}\footURL{https://huggingface.co/datasets/Akila/ForgottenRealmsWikiDataset} by~\citet{peiris2022synthesis} is of great relevance to our work. 
It focuses on the creation and evaluation of a large domain specific data set for \dnd. 
The authors describe the process of collecting, cleaning, and synthesizing data from the Forgotten Realms wikia\footnoteref{forgottenrealmswikiref} which had~\oldarticles{} at the time of the paper's publication. 
The authors use this Wikia from Fandom, Inc.\footURL{https://www.fandom.com} to create a comprehensive data set for use in various domain specific Natural Language Processing tasks~\cite{chen-etal-2021-scixgen-scientific,gu2021domain,amin-nejad-etal-2020-exploring,ferrari2017detecting} on \dnd{} domain. 
The data set is composed of multiple sub-data sets catering to different NLP tasks and needs~\cite{zhang2019bertscore,de-silva-dou-2021-semantic,sugathadasa2018legal}.

% \begin{figure}[htbp]
% \includegraphics[width=\linewidth]{images/tiamat.png}
% \caption{Screenshot: Forgotten Realms Wiki article on Tiamat}
% \label{fig:tiamat}
% \end{figure}

The authors have also evaluated semantic similarity scores on multiple metrics based on the hierarchy of first links. The paper defines first links as ``[the] first internal reference link (refers to another article in the same wikia) found in an article that is not a broken link or a miscellaneous link such as the pronunciation guide''~\cite{peiris2022synthesis}. 

Although, this is still an excellent method of automatically extracting a hierarchical structure from the data, we believe that things could be improved with a manually annotated data set. To give an example as to why that is, we will consider the article on Tiamat\footURL{https://forgottenrealms.fandom.com/wiki/Tiamat} in which, the first paragraph or the lead section is shown in Fig.~\ref{fig:tiamat_lead}. In the paragraph, the internal links in order are, \textit{lawful evil}, \textit{dragon}, \textit{evil dragons}, \textit{greater gods}, \textit{Bane}, \textit{Asmodeus}, \textit{Faerûnian pantheon}, \textit{Draconic pantheon}, and \textit{Untheric pantheon}. If we were to simply select the first link, it will point to the article \textit{Lawful evil}, when in fact the defining trait of the subject of this article is either \textit{dragon}, \textit{evil dragon} or even \textit{goddess}/ \textit{god} which is not even properly linked. This type of information extraction can be improved with manual annotation.

\begin{figure}[!htb]
    \centering
    \begin{framed}
    %\begin{quote}
    \justifying
    \scriptsize
Tiamat  was the \underline{lawful evil} \underline{dragon} goddess of greed, queen of \underline{evil dragons} and, for a time, reluctant servant of the \underline{greater gods} \underline{Bane} and later \underline{Asmodeus}. Before entering the \underline{Faerûnian pantheon}, she was a member of the \underline{Draconic pantheon}, and for some time she was also a member of the \underline{Untheric pantheon}.
\end{framed}
    \caption{Tiamat: Lead section (links are underlined)}
    \label{fig:tiamat_lead}    
\end{figure}

\section{Wikipedia lead section}
\label{subsec:lead_sec}
In a Wikipedia article, one of the most important and information-rich portion is the lead section. This refers to the first paragraph or sometimes more which appears at the top of the page before the table of contents. Compared to the rest of the content found in a Wikipedia\footURL{https://en.wikipedia.org} or any Wikipedia-esque website (that uses the MediaWiki stack and guidelines) article, this section has it's own set of specific guidelines\footURL{https://en.wikipedia.org/wiki/Wikipedia:Manual\_of\_Style/Lead\_section}. According to these guidelines, the lead section should be a summary for the entire article. And it should try to place the subject matter of the article in context with other concepts, preferably by linking the articles dedicated to the said higher concepts. 

As a result of that, typically the first link in a wiki article dedicated to a particular topic points to another wiki article dedicated to a higher concept/ broader category of the said topic. Iteratively traversed, this first link traversal can lead to higher and higher (more abstract) concepts as the articles that follow the guidelines try to put each of the topics in context. When first link traversal paths are extracted from the entire Wikia, it will form a graph (or multiple disjoint graphs) depicting the hierarchy of all the topics in the Wikia. When Wikipedia itself is concerned, about 97\% of first link traversals lead to a cycle containing the page \textit{Philosophy}\footURL{https://en.wikipedia.org/wiki/Philosophy}. In comparison, only around 30\% of first link traversals in Forgotten Realms wiki lead to a single article/ a specific traversal cycle~\cite{peiris2022synthesis}. This may be caused due to the first links in articles pointing to an article dedicated to an associate, yet not directly higher more abstract concept.

\section{Methodology}
The lexical resource for \shade{} as mentioned before, was the Forgotten Realms Wikia\footURL{https://forgottenrealms.fandom.com}. Although as part of the MediaWiki stack there is an option to export the pages in the Wiki\footURL{https://forgottenrealms.fandom.com/wiki/Special:Export}, this feature is limited when exporting a large number of pages. Due to this, we had to invoke the API endpoint of the same to export the articles. The exports are in XML format containing tags including \textit{title}, \textit{redirect}, \textit{revision}, and \textit{text} for a given article. The \textit{text} tag contains the content of the page. The MediaWiki stack uses \textit{Markdown text} as the formatting option for the article contents. So the content in the text tag contains Markdown text formatting. 

To enforce IAA and reduce human error, we propose an annotation application (\shade) that lists out the possible annotation tags/ labels for the annotators to choose from. The labels are all selected from the first paragraph (lead section) due to its property where it sets the context of the subject under discussion in the said article by explaining it with broader subject matter, in most cases referring/ linking the articles on the broader subject matter itself. This way we can minimize the suggestions to a list of highly probable labels. If for some reason the lead section is missing, we take the first paragraph under the first section and extract the labels from that. 
The annotation labels that the system provides are split into 2 different Lists. The first list contains internal links. i.e. links that refer to other articles form the same Wikia. The second list contains noun phrases extracted from the same first paragraph.

\begin{figure*}[!htbp]
\includegraphics[width=\linewidth]{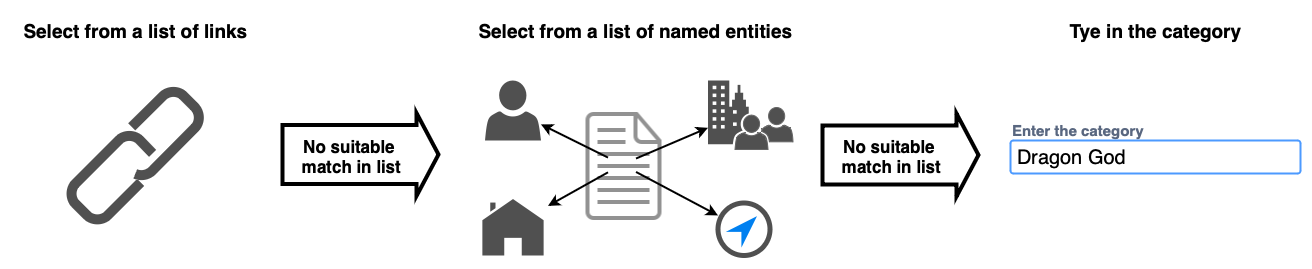}
\caption{Annotation workflow}
\label{fig:workflow}
\end{figure*}

\subsection{Escaping the infobox}
To extract the internal links in the first paragraph to populate the first list, we need some additional prepossessing on the Markdown text. Namely, we need to isolate the first paragraph in order to do so. For this, we need to remove all preceding and proceeding content. Once the beginning of the first paragraph is identified or preceding content has been removed, isolating the first paragraph by removing the proceeding text is easy enough by splitting at the first carriage return (\verb|\n|). Removing the preceding content however, can be tricky as typically, this includes the structures; infoboxes\footURL{https://en.wikipedia.org/wiki/Help:Infobox} and \verb|DEFAULTSORT|. Infobox is the table-like \textit{structure} that can be found at the top right corner of almost all Wikipedia or Wikia articles. Although in any other context, this is one of the most valuable resources, for this task, it is considered as unwanted content. The infobox is contained between double curly braces (i.e. between \verb|{{| and \verb|}}|). Apart from plain text content, infoboxes also contain links. Infoboxes and \verb|DEFAULTSORT| are typically at the very top of the Markdown content.
These structures are wrapped in two sets of curly braces. In any other scenario, to remove such structures an existing MediaWiki parser such as the \verb|MWParserFromHell| can be used, It is not applicable for this situation since we would lose the link information along with the structures leaving only the plain text. 

Due to the diverse nature of infoboxes it can include other types of structures including grids. These structures are also typically wrapped with the same double curly brace format, making complex nested structures in the Markdown code. Hence, using regex to identify and remove these structures is not a viable solution. For this we devised a simple algorithm shown in Code~\ref{algorithm}. We iterate through the text counting the numbers of double open curly braces \code{opening\_curls} and the double closing curly braces \code{closing\_curls} until we encounter a link and \code{opening\_curls~==~closing\_curls}. This means that all the structures that have opened till that point have all been closed and we have encountered a link outside of the structures. This way we can find the first instance of a link outside of a structure. And using the \code{last\_curl\_index} variable, we can also keep track of where the structures end.

\begin{lstlisting}[caption={Algorithm to remove structures},label=algorithm]
Escape_Infobox(str)
  opening_curls = 0
  closing_curls = 0
  last_curl_index = 0
  For i = 1 To str.length()
    If str[i-1] == '{' And str[i] == '{'
      opening_curls = opening_curls + 1
    If str[i-1] == '}' And str[i] == '}'
      closing_curls = closing_curls + 1
      last_curl_index = i
       
    If str[i-1] == '[' And str[i] == '[' 
     And opening_curls==closing_curls
        return str[i-1:], last_curl_index
    return str, last_curl_index
\end{lstlisting}

\subsection{Internal links list}
\begin{figure}[!htb]
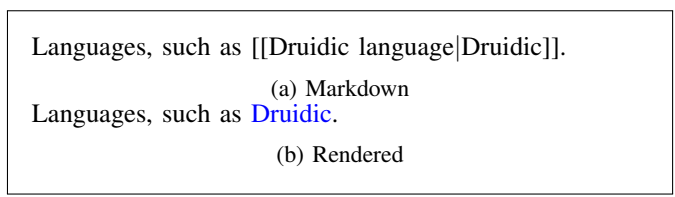

    \centering
    \begin{framed}
    \begin{subfigure}{1\textwidth}
    Languages, such as [[Druidic language$\vert$Druidic]].
    \caption{Markdown}
    \end{subfigure}
    \begin{subfigure}{1\textwidth}
    Languages, such as \textcolor{blue}{Druidic}.
    \caption{Rendered}
    \end{subfigure}
     \end{framed}
    \caption{Example link. Markdown text vs rendered version.}
    \label{fig:links}
   
\end{figure}

Internal links in MediaWiki format are included between double square braces (i.e. between \textbf{[[} and \textbf{]]}). In some cases, there is a vertical line character \textbf{$\vert$} or a ``pipe'' separating the content in the link annotation into 2 sections. The first half is the reference page title, the second is the text that will appear in the rendered version in place of the referred page title. This is what will remain when using a MediaWiki parser. Figure~\ref{fig:links} shows how a piped link appears in Markdown text and in the rendered version. 

Once the text items wrapped in double square braces have been extracted, the next step involves extracting only the reference title in cases where the link is ``piped''. In such cases, we split the text on the $\vert$ character and retrieve only the first half. This list is then saved in the database along with a weight of 1.

\begin{figure}[htbp]
\includegraphics[width=\linewidth]{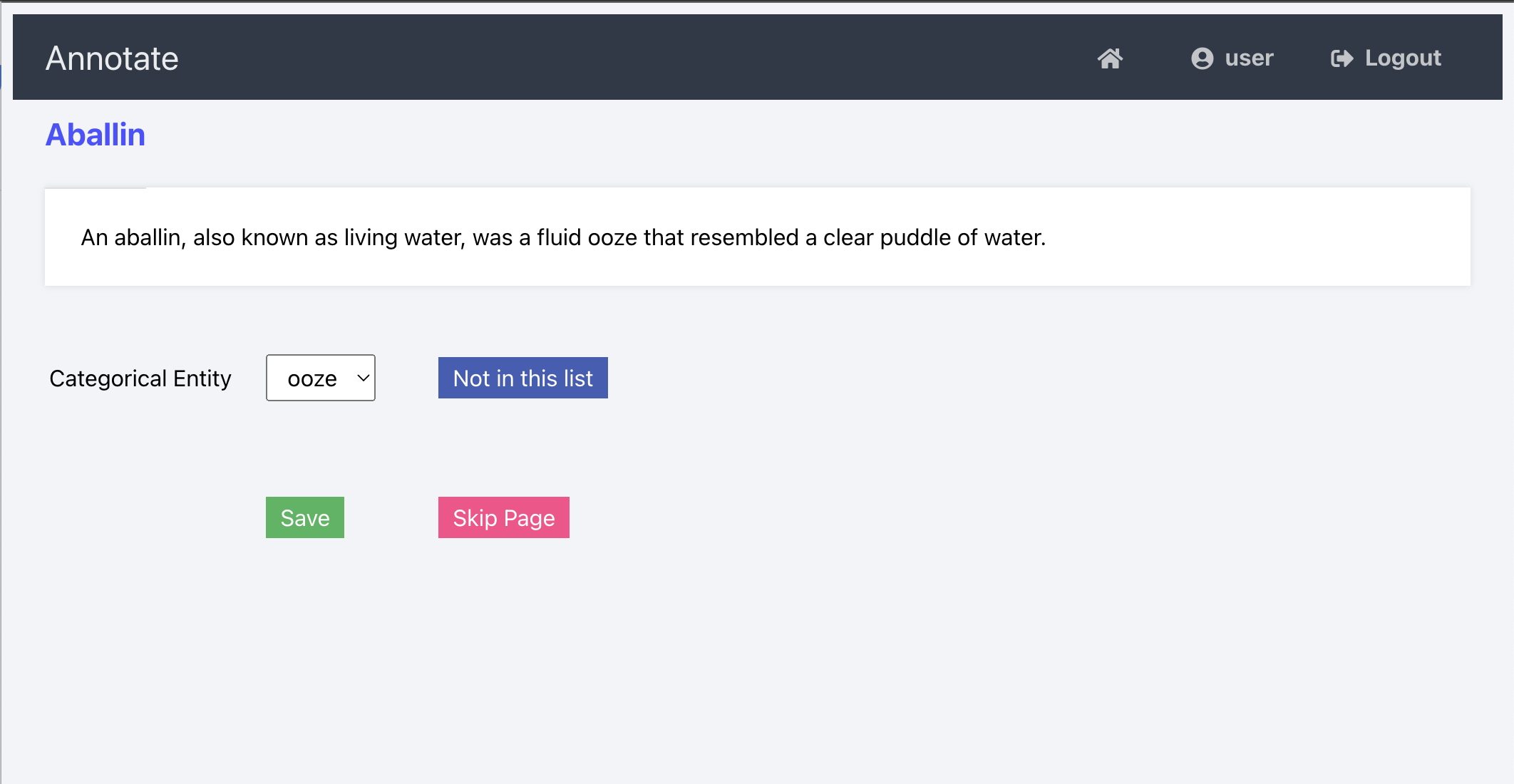}
\caption{Fist link list containing correct label}
\label{fig:screenshot1}
\end{figure}

This method does not always guarantee that all the articles will produce results. The statistics in~\cite{peiris2022synthesis} show that not all of the pages have links and even fewer has what can be considered a first link in the Forgotten Realms Wiki. And our algorithm will not consider any links that are found after the first carriage return, hence there is a possibility that the lead section spans multiple paragraphs. For this there is a base assumption and a fail-safe mechanism. The base assumption is that the lead section adheres to the lead section guidelines discussed above. In such a case, the most important links would be at the forefront explaining the subject matter with reference to the more generalized concepts. The fail-safe mechanism is the second list itself. If the first link list does not contain the most suitable label, the annotator can use the ``Not in this list'' button (blue colored button in Fig.~\ref{fig:screenshot1}) to populate the next list.

\subsection{Noun phrases list}
The second list consists of noun phrases extracted from the first paragraph/ lead section. The rationale for choosing this is that even though the pages are not linked properly, the text would reflect the intended higher classification. The article on Tiamat shown in Fig.~\ref{fig:tiamat_lead} is a good example for a cases such as this. In this example, the word ``goddess'' can be linked to reflect the article on Deity\footURL{https://forgottenrealms.fandom.com/wiki/Deity} but it has no such link configured. On the other hand, if a human annotator were to look at this, they could tag this to an article that reflects God or deity. 

\begin{figure}[htbp]
\includegraphics[width=\linewidth]{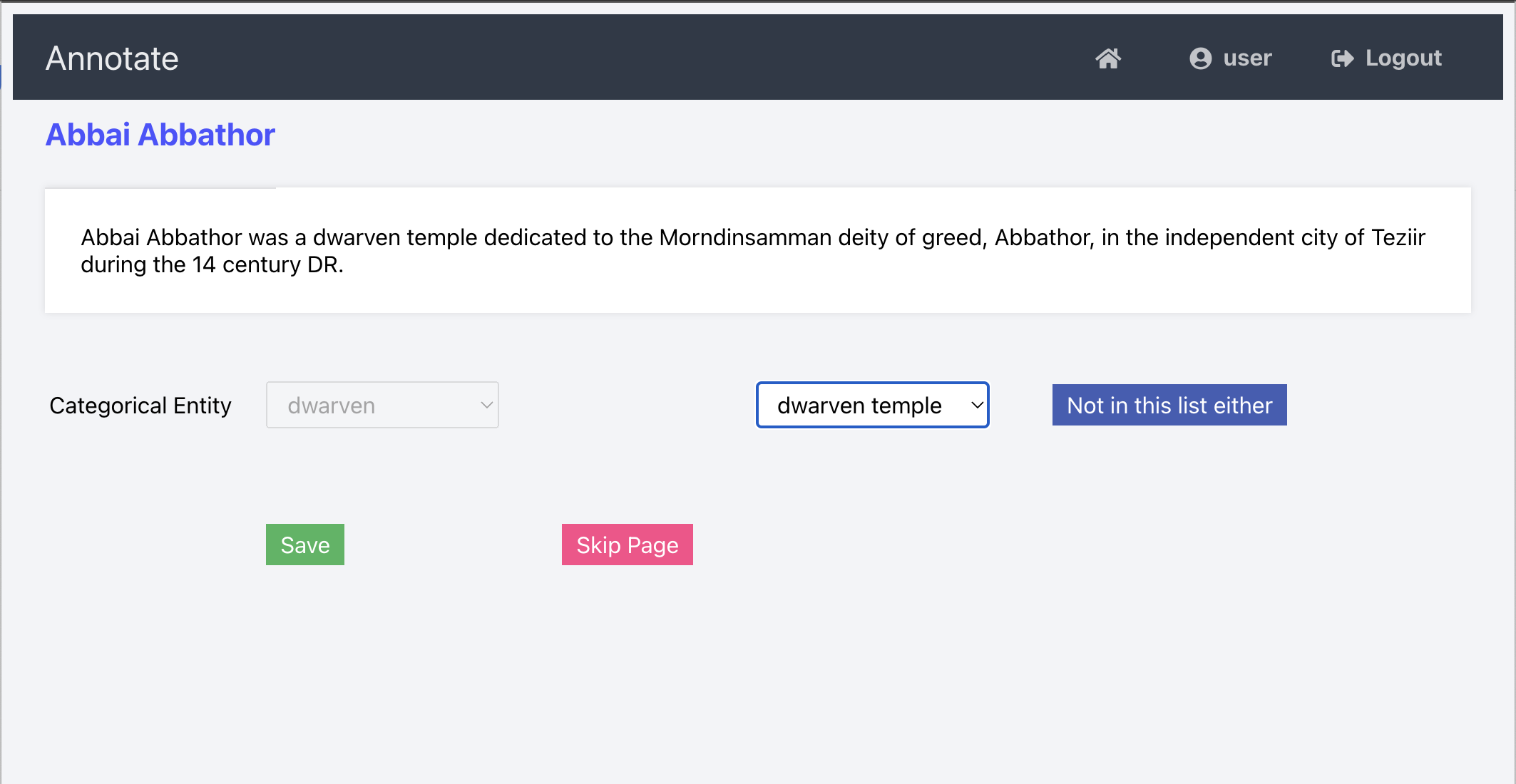}
\caption{Noun phrase list containing correct label}
\label{fig:screenshot2}
\end{figure}

Due to the fact that the  entities are described using nouns and noun phrases, by extracting these, we can create a list of potential labels that can act as a second alternative for the entity labels. To extract the noun phrases, we first extract the first paragraph. This is done so using the \code{last\_curl\_index}. The algorithm returns the last index that it had observed a curly brace in the text before encountering the first internal link outside of a structure. Here we make the assumption that the next segment is the first paragraph. Once this starting point is identified, we split the text using the index and then to remove the Markdown annotations, we use the \verb|MWParserFromHell| Python library. Once the text has been cleaned, we split it yet again in the same way as when processing the link list; at the first carriage return. The same assumptions apply here. Then to extract the noun phrase list, we use the \verb|TextBlob| Python library with a supplementary corpora. The list of extracted labels are given a weight of 2 according to the scale discussed previously. Same as with the first list, there is a dedicated button to indicate that the label is not found in this list. Figure~\ref{fig:screenshot2} showcase a situation where the correct label is found in the second list.

\subsection{Manual input labels}
This is the last failsafe mechanism that we use to capture the entity labels. By allowing the annotators to manually enter the labels, we can ensure that the probability of annotating the correct label for an entity while using the app is never 0. If the first 2 lists somehow failed to provide the correct label or the complete label (For example, in the entity \textit{Aarakocra}, the correct annotation would be \textit{avian humanoid}. But the lists suggest \textit{humanoid} only), the annotator can manually define the correct and complete label. These labels are captured with a weight of 3. 

\begin{figure}[htbp]
\includegraphics[width=\linewidth]{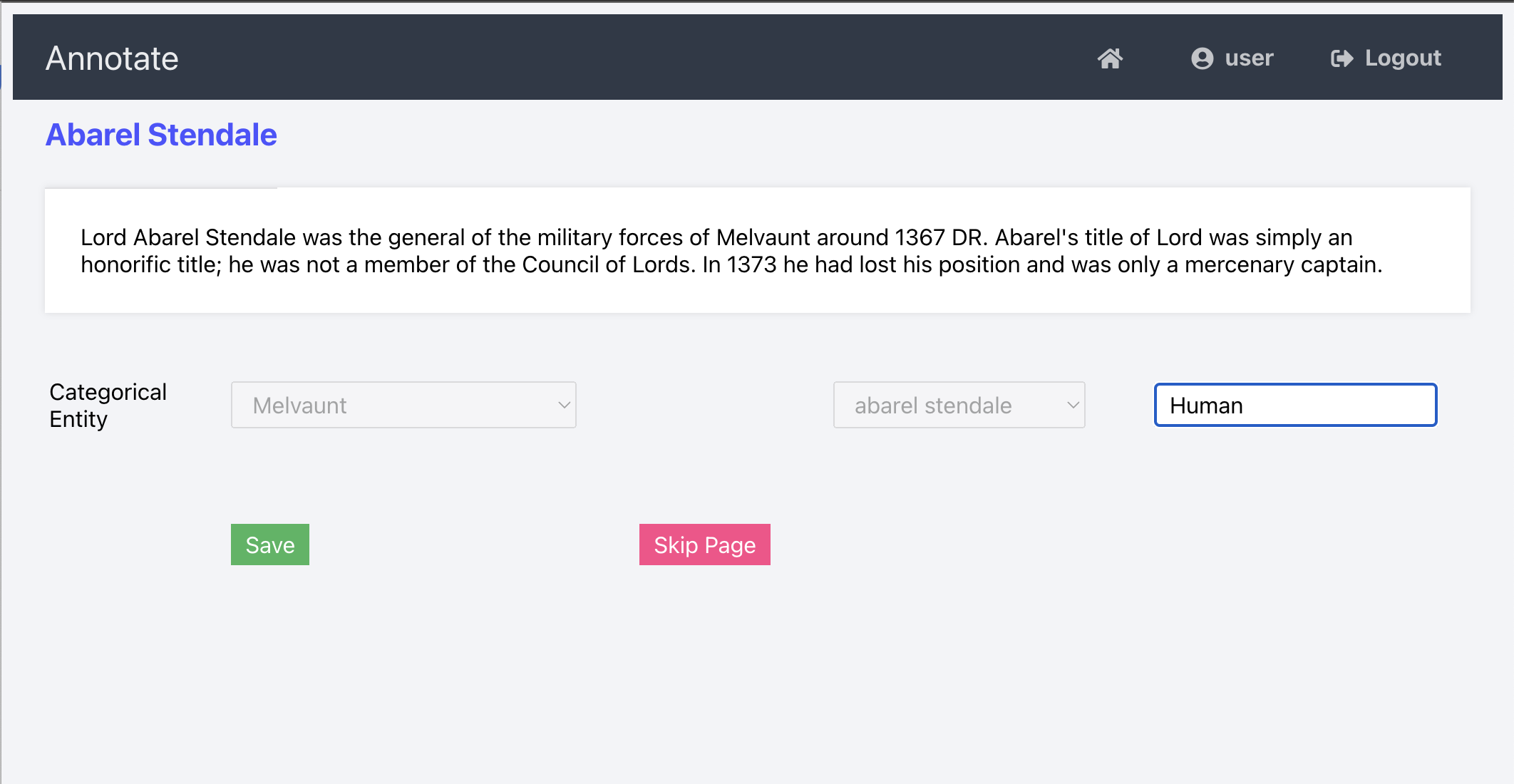}
\caption{Manually entering the correct label}
\label{fig:screenshot3}
\end{figure}

The manual input field in unlocked only when the annotator has confirmed that both of the lists do not contain the correct label as shown in Fig.~\ref{fig:screenshot3}. If the annotator needs to go back to the previous list, they can do so by refreshing the page. Once an annotator has been assigned an entity, they will always be redirected to the same entity until they complete the annotation for the entity. If the annotator is unsure with the annotation of a particular entity, they are given the option to skip that entity as well.

\begin{figure}[htbp]
\includegraphics[width=\linewidth]{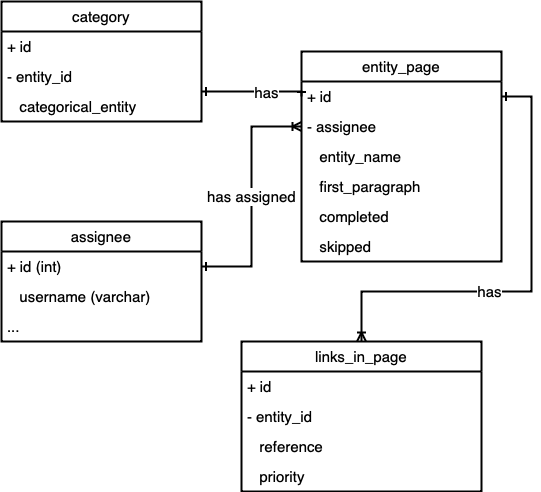}
\caption{ER diagram}
\label{fig:er}
\end{figure}

\section{Assigning and keeping track of annotations}

Figure~\ref{fig:er} depicts the entity relationship diagram for the database which contains a list of entities (\code{entity\_page}). 
Aside from the \code{entity\_name} and the \code{first\_paragraph} extracted from the wiki, this also maintains a \code{m:1} mapping with the annotators (\code{assignee}) and several states including whether the annotation was \code{completed} or was marked as \code{skipped}. By separating the completed and skipped flags, we can ensure that we can identify the entities that the annotators have deliberately marked with ``Skip Page'' button (Figure~\ref{fig:screenshot3}) from unintentionally skipped entities, for example by way of reloading the page. 

To make sure that the entities are not skipped unintentionally, we have devised a mechanism where it will present the user with the top most result that has been assigned to them (typically one result), and has the \code{completed} and \code{skipped} flags in False state. This way the annotator cannot skip a given entity unless explicitly declaring their intention to do so. By filtering out the entities where $entity_page.skipped==TRUE$, we can identify the entities the annotators have trouble labeling.

\section{Discussion}
\begin{figure}[htbp]
\centering
\includegraphics[width=\linewidth]{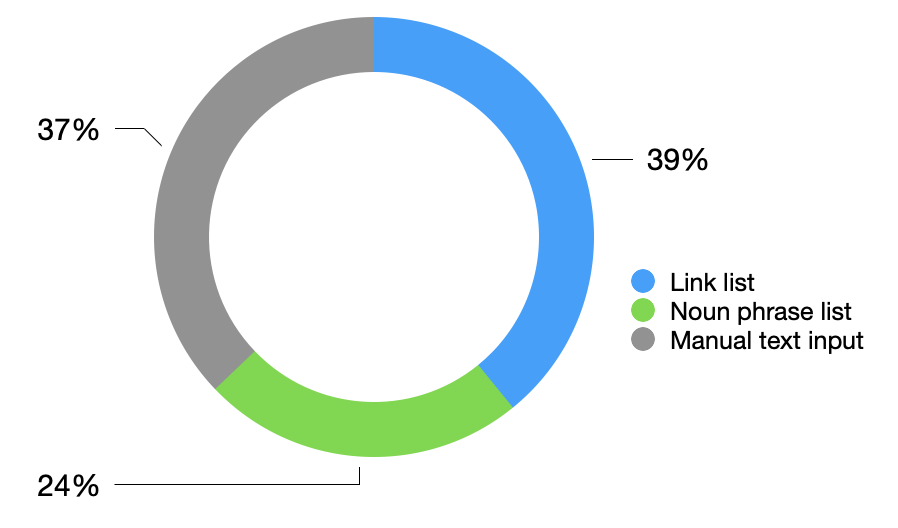}
\caption{Current annotation breakdown by label source}
\label{fig:graph}
\end{figure}

We have employed annotators to annotate the \dnd{} entities using the \shade{} application. We have currently annotated a total of \annotationsdone{} where \yearsannotated{} refer to articles on the fictional timeline. \othersannotated{} are named entities that refer to other types of concepts. Out of the \othersannotated, \listoneannotations{} were picked from the links list, \listtwoannotations{} were picked from noun phrases list, and \typedannotations{} were typed in. Figure~\ref{fig:graph} showcases the breakdown of the currently annotated entities by the label source. The two lists providing nearly 2/3 of the labels prove that this system can can reduce the manual text inputs during annotation significantly. It also provides evidence that not all the correct annotations are properly linked in the wikia, so an automatically extracted data set such as the \FRWFL{} sub data set under \FRW~\cite{peiris2022synthesis} can be enriched via manual annotation.

\section{Conclusion and Future work}

This project provides a glimpse on how we can structure an application that is used for data annotations, especially when there are text corpi associated with the entities to be annotated. Although this was designed for a specific Fandom Wikia, due to the MediaWiki guidelines, we can not only port this to other Fandom Wikis, but also any Wiki-esque data source that uses the MediaWiki stack, including Wikipedia itself. Further improvements to the \shade{} system includes allowing multiple annotations to the same entity with different priorities, and including \textit{has-a} relationship annotations.

% \section*{Acknowledgment}

\bibliography{custom}
\bibliographystyle{IEEEtranN}

\end{document}